\documentclass[conference]{IEEEtran}
\IEEEoverridecommandlockouts
\usepackage{cite}
\usepackage{amsmath,amssymb,amsfonts}
\usepackage{algorithmic}
\usepackage{graphicx}
\usepackage{subfig}
\usepackage{wrapfig}
\usepackage{pgf-pie}
\usepackage{textcomp}
\usepackage{xcolor}
\usepackage{listings}
\usepackage{multicol}
\usepackage[inline]{enumitem}
\usepackage{multirow}
\usepackage{hyperref}

\def\BibTeX{{\rm B\kern-.05em{\sc i\kern-.025em b}\kern-.08em
    T\kern-.1667em\lower.7ex\hbox{E}\kern-.125emX}}
\begin{document}

\title{Code Vectorization and Sequence of Accesses Strategies for Monolith Microservices Identification}

\author{\IEEEauthorblockN{Vasco Faria and António Rito Silva}
\IEEEauthorblockA{INESC-ID, Instituto Superior Técnico, University of Lisbon -- Lisbon, Portugal \\
\{vasco.faria, rito.silva\}@tecnico.ulisboa.pt}
}

\maketitle
\thispagestyle{plain}
\pagestyle{plain}
\begin{abstract}

Migrating a monolith application into a microservices architecture can benefit from automation methods, which speed up the migration and improve the decomposition results. One of the current approaches that guide software architects on the migration is to group monolith domain entities into microservices, using the sequences of accesses of the monolith functionalities to the domain entities. In this paper, we enrich the sequence of accesses solution by applying code vectorization to the monolith, using the \textit{Code2Vec} neural network model. We apply \textit{Code2Vec} to vectorize the monolith functionalities. We propose two strategies to represent a functionality, one by aggregating its call graph methods vectors, and the other by extending the sequence of accesses approach with vectorization of the accessed entities. To evaluate these strategies, we compare the proposed strategies with the sequence of accesses strategy and an existing approach that uses class vectorization. We run all these strategies over a large set of codebases, and then compare the results of their decompositions in terms of cohesion, coupling, and complexity.
\end{abstract}


\section{Introduction}

As microservices architectures prove their value over monoliths, an increasing number of monoliths are being migrated to the microservices architecture, which provides significant benefits in terms of scalability, agile development, and maintainability. Despite these advantages, and depending on the size and complexity of a monolith codebase, this migration process can become very complex and expensive, which makes it worth the use of tools to automate some steps of the migration.

Abdellatif et al.~\cite{Abdellatif2021} present, in a survey on the modernization approaches of legacy systems, several migration approaches for the automatic identification of microservices in monolith systems. These approaches are classified by their inputs, processes, and outputs. These approaches work on a codebase's representation obtained by applying collection tools, which can be static if they rely only on the monolith source code, or dynamic if they require the monolith execution to collect data.
However, according to this study, it can be observed that the majority of the approaches for monolith migration perform a static analysis of the source code, followed by a clustering algorithm in conjunction with similarity measures, that define the distances between the elements of the monolith that will constitute the microservices.

In what concerns the collection part, the static analysis, though the mainly used technique, becomes difficult to scale because it depends on the particular programming languages and frameworks used in the monolith implementation, which requires continuous effort in the implementation of static analyzers. On the other hand, there is a report that the use of dynamic collection techniques presents some problems with the completeness of the collection and the management of a large amount of collected data~\cite{andrade2022}.

The goal of this paper is to study whether an approach that does not require a complex static analysis data collection can generate good decompositions. This would remove some bottlenecks of previous work since the collector wouldn't be restricted to a particular programming language, web development stack, and object-relational mapper.

This approach is inspired by Al-Debagy and Martinek’s work~\cite{Al-Debagy2021}, which analyzes the monolith code like a Natural Language Processing problem (\textit{NLP}). They use a neural network model called \textit{Code2Vec} for microservices identification. This model takes advantage of a method abstract syntax tree (\textit{AST}) and the lexical interpretation of its tokens to calculate a numerical vector, representing as much information about that method as possible. With this tool, they generate vectors associated with the monolith classes and measure the quality of the decomposition in terms of cohesion and coupling metrics. However, they do not analyze the monolith from the perspective of the monolith functionalities sequences of accesses, which is one of the most common approaches, e.g.~\cite{Jin19,samuel22}.

In particular, in~\cite{samuel22}, a large number of monolith codebases are used for an extensive analysis of the identification of microservices in a monolith, which uses the monolith functionalities sequences of accesses. The study analyzes the results by applying coupling, cohesion, and complexity metrics for the generated decompositions.

In this paper, we leverage on~\cite{Al-Debagy2021,samuel22}, by integrating their perspective to verify, using a larger number of codebases, whether:
\begin{enumerate}
    \item The use of \textit{Code2Vec} with the functionality perspective provides better results than sequences of accesses in~\cite{samuel22};
    \item The application of the functionality perspective provides better results than Al-Debagy and Martinek~\cite{Al-Debagy2021};
    \item The input parameters of the proposed strategies impact the results of the evaluation metrics.
\end{enumerate}

The proposed solution starts with a Data Collection phase, where a new collector is used to extract all the methods of a monolith codebase, along with all their information (package, class, type, source code, and method calls). During this phase, the \textit{Code2Vec} model is used to generate each method's respective vector. After, we test two different strategies to generate the functionalities vectors.

For evaluation, we apply the different strategies to a large set of codebases, and then compare the results using cohesion, coupling, and complexity metrics.

After this section, Section~\ref{sec:related_work} discusses work related to the application of machine learning techniques in software migration, followed by, Section~\ref{sec:background}, a short description of the background required to the proposed solution. Section~\ref{sec:solution} presents new strategies for microservices identification in monolith systems. Section~\ref{sec:evaluation} evaluates and compares the new strategies with previous work, and finally, Section~\ref{sec:conclusion} presents the final conclusions of this work.

\section{Related Work}
\label{sec:related_work}

Since the emergence of microservices architectures, migrating monoliths to these architectures has been an increasingly active topic~\cite{Abdellatif2021}.

There are approaches~\cite{Jin19,Nunes2019,samuel22} that use the monolith functionalities sequences of accesses to the monolith domain entities to feed an aggregation algorithm that proposes candidate decompositions for microservices. These approaches can use static analysis of the monolith code, e.g.~\cite{samuel22}, or dynamic execution of the monolith to collect the sequences of accesses. Andrade et al.~\cite{andrade2022} compares the use of static and dynamic collection for microservices identification. They conclude that, while in static analysis the data collection needs to be adapted to each programming language or framework, which requires tool adaption effort for each new programming language of full-stack technology, the dynamic collection of data shown to have worse coverage, though generating a huge amount of data. Based on these results, our research intends to explore the use of lexical analysis, a form of static analysis that requires less effort because it is more language and technology independent.

On the other hand, several approaches for the identification of microservices, or grouping classes into packages, apply lexical analysis.

Hammad and Banat~\cite{Hammad2021} propose a technique that utilizes the K-Means clustering algorithm~\cite{lloyd1982} to group a set of classes into packages, where the similarity measure presented consists of how many relevant tokens two classes have in common. This approach didn't achieve good results in terms of modularity, because the tokens must be identical in order to find a similarity between two parts of the code, which ignores all words that belong to the same semantic or lexical field.

Mazlami et al.~\cite{Mazlami2017} present three formal coupling strategies to generate a weighted graph from the meta-information of a monolithic codebase. They decompose the generated graph using a graph-based clustering algorithm. One of the strategies follows the same logic of the previously mentioned approach~\cite{Hammad2021}, based on coupling two classes containing the same tokens but considering their frequency. Although this strategy presents a worse execution time when compared with their other approaches, it shows better results in what concerns the team size reduction and the average domain redundancy.

Brito et al.~\cite{Brito2021} use topic modeling to identify services according to domain terms, where words with higher probabilities indicate a possible good topic. They also refer that the relevant tokens extraction could be easier with a pure Natural Language Processor (\textit{NLP}), but the results would be worse. This topic modeling approach is also agnostic of the development stack, but the results depend on an optimal lexical token extraction, using specific parsers for each language to extract and process the \textit{ASTs} as input to the model, which means increasing its complexity. As a result of their work, the model shows good cohesion values of the identified microservices, with the trade-off of generating a high number of clusters in order to achieve good values in a metric that evaluates whether microservices follow the Single Responsibility principle~\cite{martin2003}.

Nowadays, there are not many approaches besides clustering when it comes to machine learning techniques to decompose monolith applications into microservices. However, the use of \textit{NLPs} has been increasing due to their significant progress in performing lexical analyses, making the Data Collection phase easier and more generic.

Ma et al.~\cite{Ma2018} propose a solution based on Word2Vec~\cite{mikolov2013}, a widely-used machine learning method in Natural Language Processing, to match existing microservices to new requirements. Their approach only works for applications where scenarios are written in a common language describing the features of the target system and that already follow a microservices architecture since their goal is to discover where to place new requirements. They use the vectors generated by the Word2Vec model as the similarity measure between scenarios.

Leveraging on the Word2Vec~\cite{mikolov2013} work, Alon et al.~\cite{alon2019} created \textit{Code2Vec}, a neural network model trained to represent methods as fixed-length numerical vectors, also called code embeddings.

Al-Debagy and Martinek~\cite{Al-Debagy2021} propose an approach to decompose a monolith application into microservices using \textit{Code2Vec}~\cite{alon2019}, by extracting the methods' code embeddings.
Using these vectors, they define a class embedding as the aggregation of its methods' embeddings. After testing, they found that the mean is the most suitable aggregation function to define a class embedding. The results of this novel approach show high cohesion values since all the semantically similar classes are grouped in a microservice, making this solution achieve even better results than the other approaches they consider in the evaluation.

Overall, although there is some work on the use of \textit{Code2Vec} for the identification of microservices in a monolith, it does not follow an approach where the data collected from the monolith is based on the functionalities accesses to domain entities. Additionally, there is a lack of studies that compare the approaches for a large number of codebases, using different quality metrics.

\section{Background}
\label{sec:background}

\textit{Code2Vec}~\cite{alon2019} is a neural network model trained to represent methods as fixed-length numerical vectors, also called code embeddings. In machine learning, an embedding is a low-dimensional vector that represents high-dimensional data preserving the most information possible. Although the model is designed for method naming, the learned code embeddings can be used for several other applications.

The first stage of \textit{Code2Vec} consists of transforming a code snippet into abstract syntax tree (\textit{AST}) paths, since it improves scalability while training the model, avoiding the costs of learning the language syntax itself.

Following the extraction of the \textit{AST} paths, each path is mapped into a three-value tuple composed of the path's start node, intermediate expressions, and the final node. Then, each part of the tuple is converted to a real-valued representation, creating a three-dimension numerical vector known as a context vector acting as input to the path-attention network.

A neural attention network architecture is used to overcome the data sparsity problem of similar methods having different \textit{ASTs} paths. With this attention mechanism, the model also learns the importance of each path, applying higher weights to the most important ones.

By applying the learned weights and the hyperbolic tangent function on the input vectors, the code embedding is computed using the attention weights to calculate a weighted average of all the combined context vectors.

\section{Solution}
\label{sec:solution}

\subsection{Data Collection}

The first step of our approach consists of extracting all the necessary information from the monolith codebase and preparing it. This is done using the JavaParser library\footnote{https://javaparser.org/}, which is a popular static analysis tool used to parse and modify java code by generating an interactive abstract syntax tree and providing a symbol resolution module. JavaParser also provides a type resolution module (symbol-solver) that can combine different type solvers to increase the capability of solving complex references like superclass methods.

For the data collection, we explore all the codebase files, recurring to the JavaParser type solvers. For each java file, our parser starts by identifying the package, the class/interface name, the annotations, and all the present methods as well as checking if the class extends another.

Every time the parser founds a new method, its respective body is converted into a code embedding by the \textit{Code2Vec} model, which we save along with the method signature. Also, inside each method body, the parser looks for all methods invocations' and tries to solve their signature using the type solvers. If the invoked method belongs to external libraries of the codebase, those invocations are discarded.

Since the evaluation is applied to monoliths implemented using Spring-Boot and an Object-Relational Mapper (\textit{ORM}) each code embedding is characterized in terms of Spring-Boot architectural elements: Controller, Entity, Service, Repository, and Configuration classes. This categorization is used to verify whether some parts of the monolith code can provide more accurate results, and to identify the starting point of each functionality (Controller) and what are the monolith persistent domain entities (Entity).

\subsection{Functionality Vectorization Strategies}

We propose two functionality vectorization strategies to represent a functionality as an embedding by using the functionality call graph, or the functionality sequence of accesses to domain entities. The purpose of these strategies is to represent each microservice as a set of functionalities and thus understand which functionalities should be implemented in the same microservice.

\begin{figure}[ht]
\centering
\includegraphics[width=7cm]{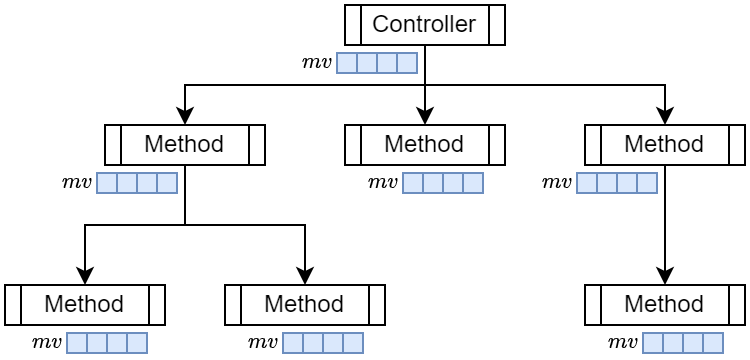}
\caption{Extraction of a functionality call graph vector.}
\label{fig:cgv}
\end{figure}

Figure~\ref{fig:cgv} presents the \textbf{Functionality Vectorization by Call Graph} (\textit{FVCG}) strategy, which represents each functionality as the call graph of its methods invocations, where the first method is the controller where the functionality starts executing. By traversing a method call graph it is possible to reach loops, so to overcome this problem, a maximum depth parameter on the call graph is considered to compute the vector.

After discovering all the methods and the respective code embeddings, represented in Figure~\ref{fig:cgv} by the $mv$ vectors, that belong to the call graph of a functionality for a given depth, we apply the mean weighted function to those embeddings in order to achieve the functionality representing embedding. The method annotations are used to infer each method type. The following weights for method types are considered:

\begin{itemize}
    \item \emph{$w_c$}: The controllers weight;
    \item \emph{$w_s$}: The services weight;
    \item \emph{$w_e$}: The entities weight;
    \item \emph{$w_i$}: The remaining methods (e.g., auxiliary or unclassified methods) weight, which will be referred as intermediate.
\end{itemize}

The weights are positive values that should sum 100 ($w_c + w_s + w_e + w_i = 100$).

\begin{figure*}
    \centering
\begin{equation}
cgv(f) = \frac
{
    \sum_{mv \in f.cg(d).C} w_c \times mv +
    \sum_{mv \in f.cg(d).S} w_s \times mv +
    \sum_{mv \in f.cg(d).I} w_i \times mv +
    \sum_{mv \in f.cg(d).E} w_e \times mv
}
{
    \sum_{mv \in f.cg(d).C} w_c +
    \sum_{mv \in f.cg(d).S} w_s +
    \sum_{mv \in f.cg(d).I} w_i +
    \sum_{mv \in f.cg(d).E} w_e
}
\label{eq:cgv}
\end{equation}
\end{figure*}

The vector is computed according to equation~\ref{eq:cgv}, where $f.cg(d)$ denotes the functionality ($f$) call graph, generated with depth $d$, due to possible recursive invocations, and $.C, .S, .I, .E$, denote, the call graph nodes that are of type, respectively, controller, service, intermediate and entity. Note that, additionally to the weight parameters, $d$ parameter on the call graph depth determines the number of method vectors to consider. The purpose of these parameters is to study their impact on the quality of the result, and how they affect the evaluation metrics results. This study will help to understand the level of computational effort required in the construction of the vectors. For instance, if vectors computed using low depth provide good results, it will significantly reduce the computational effort. On the other hand, if the weights are irrelevant, the data collector will not need to recognize the type of each method, being a positive aspect to make the collector framework and architecture agnostic.

\begin{figure}[ht]
\centering
\includegraphics[width=7cm]{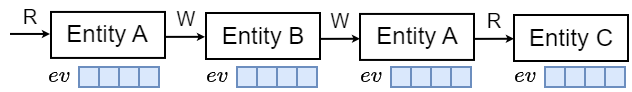}
\caption{Extraction of a functionality sequence of accesses vector, where \textit{R} stands for read and \textit{W} stands for write.}
\label{fig:sv}
\end{figure}

Figure~\ref{fig:sv} presents the \textbf{Functionality Vectorization by Sequences of Accesses} (\textit{FVSA}) strategy, which represents each functionality as the sequence of its accesses to domain entities, where read accesses are distinguished from write accesses. It uses the sequences of accesses done by a functionality, and associates to each access the embedded vector of the accesses entity, $ev$.  The entity embedded vector is computed by first identifying the entity methods, and then calculating the mean of that method's embeddings. 

In order to represent entities as the mean of its methods' embeddings all classes should have methods. This is not the case of some classes that extend other classes or use annotations to generate their methods in compile time. To mitigate this problem, we include inheritance in each class embedding by using all the top hierarchy classes' methods in the aggregation function. However, the case of methods generated through annotations at compile time, are not addressed and their empty classes are not considered. In the experiment, codebases having these type of classes are excluded.

\begin{equation}
sav(f) = \frac
{
    \sum_{ev \in f.sa.R} w_r \times ev +
    \sum_{ev \in f.sa.W} w_w \times ev
}
{
    \sum_{ev \in f.sa.R} w_r +
    \sum_{ev \in f.sa.W} w_w
}
\label{eq:sv}
\end{equation}

Having functionality sequences of accesses and the entities' embeddings, the functionality embedding is the weighted average of the entities' embeddings of all the entities possibly accessed during the functionality execution, as presented in equation~\ref{eq:sv}. In the equation, the entities read by functionality $f$ in its sequence of accesses are denoted by $f.sa.R$, while $f.sa.W$ denotes the entities written. The parameters $w_r$ and $w_w$, represent, respectively, the weight associated with the type of access, read and write. The weight values are positive and should sum to 100. Note that, as in the previous vectorization, the parameters will be used to assess the impact of distinguishing reads from write accesses in the quality of the generated decompositions.

\section{Evaluation}
\label{sec:evaluation}

To answer the research questions, we compare the \textit{Code2Vec} decompositions generated using the \textit{Code2Vec} similarity measures built on the monolith functionalities with the decompositions generated using sequences of access, as in~\cite{Santos2020}, the decompositions generated using vectors for classes built with \textit{Code2Vec}, as in~\cite{Al-Debagy2021}, and the decompositions that only consider vectors for entities, which is a sub-category of the previous strategy~\cite{Al-Debagy2021}.

\subsection{Strategy Comparison}

The strategy by Al-Debagy and Martinek~\cite{Al-Debagy2021} represents a microservice as a set of classes. They use \textbf{Class Vectorization} (\textit{CV}), where each class has an embedding calculated as the mean of its methods embeddings, a method already applied in the \textit{FVSA} strategy.

Nevertheless, there are approaches where microservices are represented by monolith domain entities, instead of their classes, to highlight that the main aspect of a microservice is the independence of its database from other microservices databases. Therefore, we use another strategy adapted from the \textit{CV} strategy in which, rather than representing a microservice as a set of classes, it is represented as a set of entities. The \textbf{Entity Vectorization} (\textit{EV}) strategy only considers the classes in \textit{CV} strategy that are entities.

The third strategy we are going to compare to is the \textbf{Sequence of Accesses} (\textit{SA}). There are four similarity measures based on the sequences of access~\cite{Santos2020}. They aggregate the monolith domain entities that are accessed by the same functionalities. The main idea behind these measures is that in a microservices architecture it is necessary to minimize the number of distributed transactions. Therefore, by having all the domain entities that are accessed by a functionality in the same cluster, the functionality can execute as a single transaction. These similarity measures represent the distance between two domain entities by using the sequences of accesses strategy (\textit{SA}) of the functionalities that access them. Therefore, each of the similarity measures between entities $e_i$ and $e_j$ are defined as the following:

\begin{enumerate}
    \item \textit{Access}: Given a set of functionalities that access, both read or write, entity $e_i$, is the percentage of those who also access entity $e_j$.
    \item \textit{Read}: Given a set of functionalities that read entity $e_i$, is the percentage of those who also read entity $e_j$.
    \item \textit{Write}: Given a set of functionalities that write entity $e_i$, is the percentage of those who also write entity $e_j$.
    \item \textit{Sequence}: The percentage of the number of consecutive accesses to $e_i$ and $e_j$ entities over the maximum number of consecutive accesses for two domain entities.
\end{enumerate}

Note that these measures, except the sequence, are not symmetric.

The \textit{SA} strategy uses the four similarity measures by assigning weights to each one of them, such that their sum should be 100.

To compare the different strategies, they have to produce the same type of decomposition clusters. However, the strategies produce three different types of decompositions. \textit{SA} and \textit{EV} strategies generate clusters of entities, \textit{FVCG} and \textit{FVSA} strategies clusters of functionalities, and the \textit{CV} strategy clusters of classes. Therefore, to compare the results, it is necessary to convert a decomposition type into the other. Since the metrics to be used in the evaluation are defined as decompositions of clusters of domain entities, the decompositions are converted into decompositions with clusters of entities.

To convert a cluster of classes into an entity's clusters, it is only necessary to remove all the non-entity classes from the clusters, which can lead to empty clusters and so we discard those clusters.

The functionalities clusters are converted into clusters of entities by counting the functionalities entity accesses present in each cluster. This is, for each domain entity's access by a functionality of a given cluster, the probability of that entity belonging to that respective cluster increases. Then, for each domain entity, we look for the cluster that accesses it the most to assign the entity to that respective cluster. Afterward, since this conversion may also result in empty clusters, those are discarded.

\subsection{Decomposition Generation}

To evaluate the strategies it is necessary to generate a significant number of decompositions, varying the number of clusters and the strategies weights.
In terms of the number of clusters, for codebases up to 10 entities, a maximum of 3 microservices are generated, between 10 and 20 a maximum of 5 microservices, and for more than 20 the maximum number of microservices is 10.

We start at a minimum of 3 microservices and generate all possible decompositions by varying the strategy's parameters. Then we repeat the process by increasing by one the number of microservices to generate, until we reach the maximum number of microservices.
The strategies that don't represent a microservice by a set of entities may result in empty clusters. Therefore, the real number of clusters of the generated decompositions is smaller than the requested one. To overcome this issue, we continue to increase the requested number of microservices and generate the respective decompositions until we achieve one that results in a number of clusters bigger than the maximum value.

The number of decompositions generated for each strategy depends on the number of its parameters since we explore all the possible combinations. For the weight parameters, we need to create all the combinations where the sum of the weights equals 100, using intervals of 10. In the \textit{FVCG} strategy, we decided to vary the depth parameter from 1 to 6.

A hierarchical clustering algorithm is applied to the strategies vectors and distances, using the euclidean distance. A dendrogram is generated, which is cut to generate decompositions with different numbers of clusters. 

Since the hierarchical clustering algorithm supports different types of linkage criteria to determine the distances of the clusters, they are also used as variations in the evaluation. The three criteria considered are:

\begin{itemize}
    \item Single-linkage clustering: Distance between the closest entities of the measured clusters.
    \item Complete-linkage clustering: Distance between the furthest entities of the measured clusters.
    \item Average-linkage clustering: Average of the distances between each entity of one cluster and the entities of the other.
\end{itemize}

This linkage type parameter will also be exercised during the generation of decompositions.

\subsection{Evaluation Metrics}

Three metrics are used to evaluate the quality of a generated decomposition: coupling, cohesion, and complexity.

The cohesion measures the single responsibility principle~\cite{martin2003}. The cohesion of a decomposition is computed using the cohesion of each one of its clusters. The cluster cohesion is percentage of the cluster's entities accessed by the respective functionalities. Therefore, a cluster has higher cohesion if the accesses done by functionalities interact with all the entities in the cluster. And so, it has low cohesion if each functionality, that access the cluster, only accesses a small subset of the cluster entities.

The coupling reflects the interdependence between microservices. This is measured by the percentage of entities a cluster has to know of another. A cluster knows the entity of another cluster if there is a functionality that immediately after the access of an entity in the first cluster accesses an entity in the second cluster. For instance, there is low coupling from cluster $c1$ to cluster $c2$, if all functionalities that have an access in $c1$ and immediately and access in $c2$, always access the same entity in $c2$, and $c2$ has several other entities. The accessed entity is in $c2$ interface from the point of view of $c1$. Note that coupling is not a symmetric property because it depends on the order of accesses. On the other hand, it differs from cohesion because only the pairs of accesses where two clusters are involved are relevant. The coupling of a decomposition is the average of the coupling between all pairs of the decomposition cluster.

Complexity measures the effort required to migrate a functionality from a monolith to a microservices architecture~\cite{Santos2020}. This complexity results from the need to introduce a set of distributed transactions to implement the functionality. Since the distributed transactions execution needs to be implemented using eventual consistency, due to scalability~\cite{Fox99}, it is necessary to change the business logic to consider intermediate states of the domain entities, which is a consequence of the lack of isolation. Therefore, the complexity depends on the number of distributed transactions required to implement a functionality, and the number of intermediate states they introduce. The former is calculated by how many times the functionality sequence of accesses is split between clusters, each split is a local transaction part of the overall distributed transaction. The latter is calculated by identifying the read and write accesses done by the local transactions. The complexity of a decomposition is the sum the complexity of each one of the functionalities.

An additional metric is built combining the three metrics, to evaluate which decompositions have a better balance between them, as presented in equation~\ref{eq:decomp-coupling}. Note that, the complexity is divided by the maximum complexity of all decompositions, to obtain a value between 0 and 1, this is called uniform complexity. The cohesion has a negative value because higher cohesion is better than lower cohesion, whereas, for instance, lower coupling (complexity) is better than higher coupling (complexity).

\begin{small}
\begin{equation}
\frac
{
    1 + \frac{complexity(d)}{max\_complexity(D)} + coupling(d) - cohesion (d)
}
{
    3
}
\label{eq:decomp-coupling}
\end{equation}
\end{small}

\subsection{Codebase Sample}

To gather the codebases sample for this experiment, a list of GitHub repositories that depend on the Spring Data JPA library\footnote{https://github.com/spring-projects/spring-data-jpa/network/dependents} was filtered to exclude codebases with less than five domain entities and controller classes. After that, the remaining codebases were sorted by the number of GitHub stars and manually selected from the top in order to keep the sample quite diverse in terms of codebase sizes. From these codebases we still had to exclude a few due to the dependence on libraries that generate methods from annotations on compile time, making these methods not available for a static analysis.

\begin{figure}[ht]
\centering
\includegraphics[width=7cm]{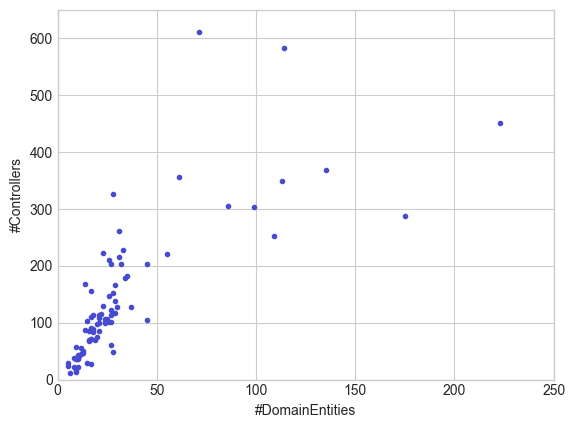}
\caption{Representation of the 85 codebases used in the evaluation.}
\label{fig:codebases}
\end{figure}

The selection process led to a relatively large number of monolith codebases (85), with an average number of code lines around 25 thousand and a standard deviation of 33 thousand lines of code, indicating a high variation of the codebases size. Also, it is possible to observe the distribution of the number of controllers and domain entities in Figure~\ref{fig:codebases}.

\subsection{Statistical Analysis}

To validate the research questions we start to compare the strategies for the cohesion, coupling, complexity, and combined metrics, using decompositions for the 85 codebases chosen for different numbers of clusters. To measure whether the differences in the results of the strategies are statistically significant, we apply the Welch's t-test~\cite{welch1947}.

Welch's t-test~\cite{welch1947} is a two-sample location test used in statistics to test the hypothesis that two populations have equal means and it is more reliable when the two samples have unequal variances and possibly unequal sample sizes, which is the case. The hypotheses of the Welch's t-test are the following:

\begin{itemize}
    \item $H_0$: $\mu_1 = \mu_2$, the samples have equal means;
    \item $H_1$: $\mu_1 \neq \mu_2$, the samples have distinct means.
\end{itemize}

To reject or accept the presented null hypotheses, we use a significance level of 0.05.

In addition, we also analyze each proposed strategy individually to study the impact of the strategy parameters on metrics values. To do so, we run regressions for each type of parameter applying the ordinary least squares (\textit{OLS}) method to choose the regression parameters, $\beta_i$ and $cons$ of the equation~\ref{eq:ols-regression}.

\begin{equation}
metric(d) = \sum_{i \in parameters}{\beta_i \times w_i} + cons
\label{eq:ols-regression}
\end{equation}

To test these regressions, we also use a significance level of 0.05 to accept or reject the following hypotheses:

\begin{itemize}
    \item $H_0$: $\beta_i = 0 \; \forall i \in \text{ }]0, \#parameters]$, the evaluation metrics do not have any relation with the parameters under analysis;
    \item $H_1$: $\beta_i \neq 0 \; \exists i \in \text{ }]0, \#parameters]$, the evaluation metrics are at least affected by one of the parameters under analysis.
\end{itemize}

For the functionality vectorization strategies, since they use parameter weights, it will be necessary to study the problem of multicollinearity, since the weights depend on each other by adding up to 100. To overcome this problem, we repeat the regression analysis without one of the dependent parameters, and by doing this for each parameter we can retrieve better coefficient values. During the evaluation, the parameters of the clustering algorithm will also be considered as parameters of the strategies to understand the impact of the linkage criteria.

\subsection{Results}

\begin{figure*}
\centering
\subfloat[Uniform complexity\label{fig:gen_complexity}]{\hspace{0.05\linewidth}\includegraphics[width=0.4\linewidth]{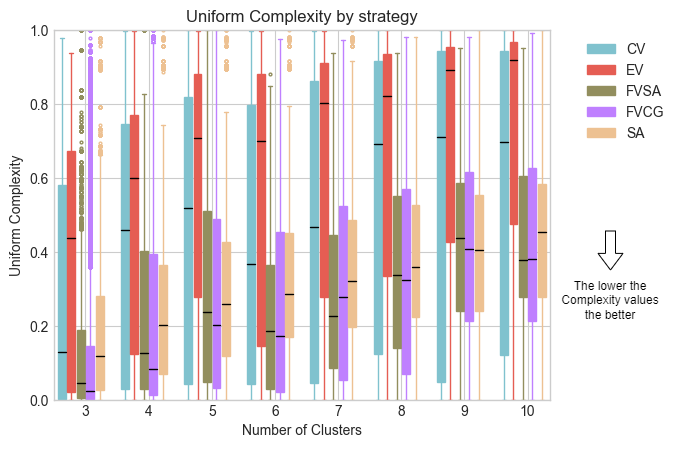}}
\subfloat[Coupling\label{fig:gen_coupling}]{\includegraphics[width=0.4\linewidth]{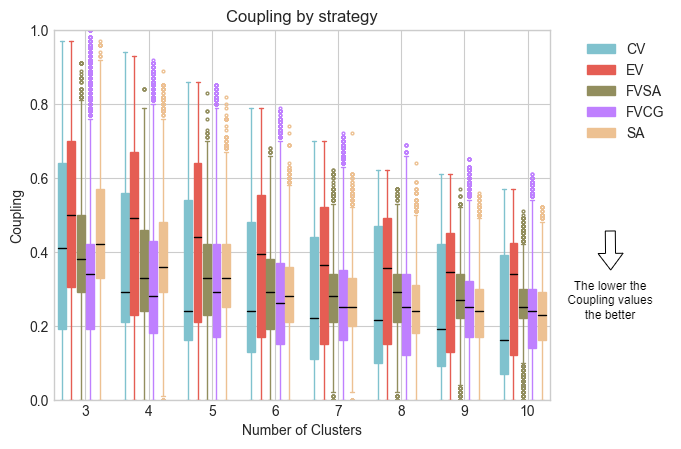}}
\newline
\subfloat[Cohesion\label{fig:gen_cohesion}]{\includegraphics[width=0.4\linewidth]{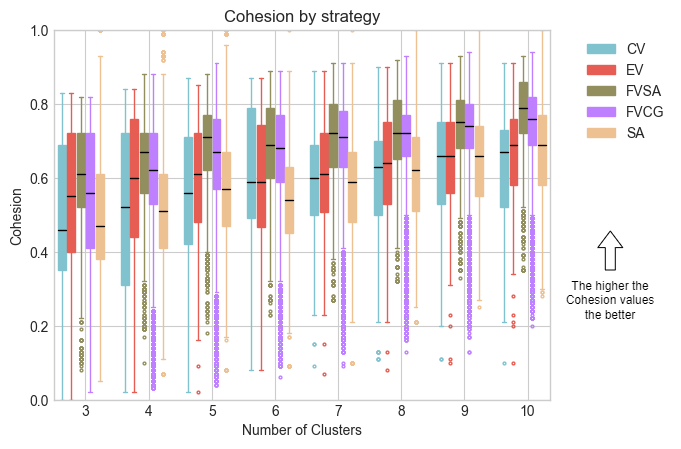}}
\subfloat[Combined\label{fig:gen_mix}]{\includegraphics[width=0.4\linewidth]{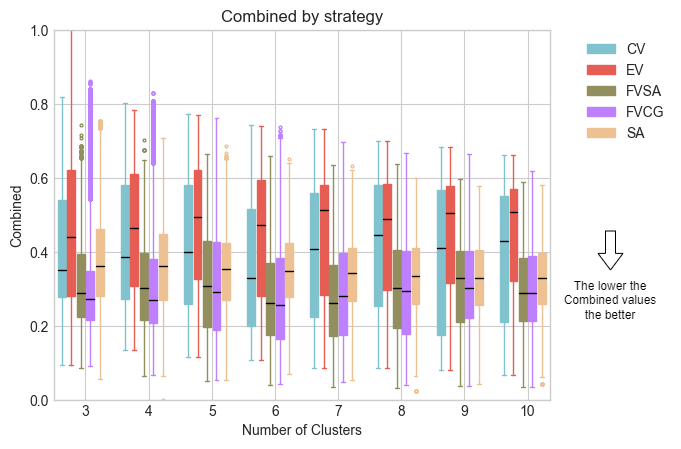}}
\caption{Evaluation Metrics applied to the 85 codebases}
\label{fig:gen_metrics}
\end{figure*}

To answer the research questions, we went through all the generated decompositions to calculate the respective values for cohesion, coupling, complexity, and combined metrics. With these values, it is possible to compare the strategies and look for any correlation between the proposed strategies parameters and the metrics values. Figure~\ref{fig:gen_metrics} presents the results.

We start by comparing the results for strategies \textit{FVSA} and \textit{FVCG}.
For complexity, the Welch's t-test rejects the hypothesis of having the same mean values except when the number of clusters is 9 and 10, and through Figure~\ref{fig:gen_complexity} it is possible to notice that most of the \textit{FVCG} strategy values are most of the times lower than those of the \textit{FVSA}, since the median is lower. Thus, it can be concluded that the decompositions generated by the \textit{FVCG} strategy are in general less complex than those generated by the \textit{FVSA} strategy.

Considering coupling, the hypothesis of having the same median values in each number of clusters is also rejected by Welch's t-test. From Figure~\ref{fig:gen_coupling} it is possible to observe that the coupling values for the \textit{FVCG} strategy are lower than those of the \textit{FVSA} strategy.

As for cohesion, (Figure~\ref{fig:gen_cohesion}), the \textit{FVSA} seems to obtain best results than \textit{FVCG} since Welch's t-test rejects the hypothesis of having the same cohesion mean values and most of those results are higher than the ones generated by the \textit{FVCG}.

In addition to these metrics, it is interesting to analyze the combined metric in figure~Figure~\ref{fig:gen_mix} that represents the balance between the previous ones. Welch's t-test only accepts the hypothesis of both strategies have the same mean when the number of clusters is 6 and 9, and as the values of the \textit{FVCG} strategies are lower than the ones of \textit{FVSA}, except for 7 and 10 cluster, the \textit{FVCG} achieve the best-balanced results.

\begin{table}[ht]
\caption{Average number of decompositions and duration of each strategy when generating all decompositions by permuting the strategy parameters}
\label{tab:strategies-performance}
{
\begin{tabular}{|l|l|l|l|l|l|}
\hline
 &  \textit{CV} & \textit{FVCG} & \textit{FVSA} & \textit{SA} & \textit{EV}\\
\hline
\#Decompositions Mean & 71 & 131168 & 503 & 1514 & 17\\
Performance Time Mean (s)  & 62  & 2830 & 350 & 39 & 32\\
\hline
\end{tabular}}
\end{table}

\begin{figure*}
\centering
\subfloat[Uniform complexity\label{fig:best_gen_complexity}]{\hspace{0.05\linewidth}\includegraphics[width=0.4\linewidth]{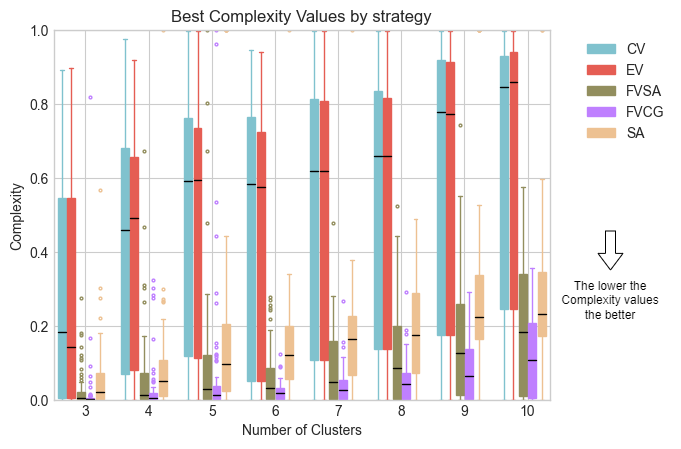}}
\subfloat[Coupling\label{fig:best_gen_coupling}]{\includegraphics[width=0.4\linewidth]{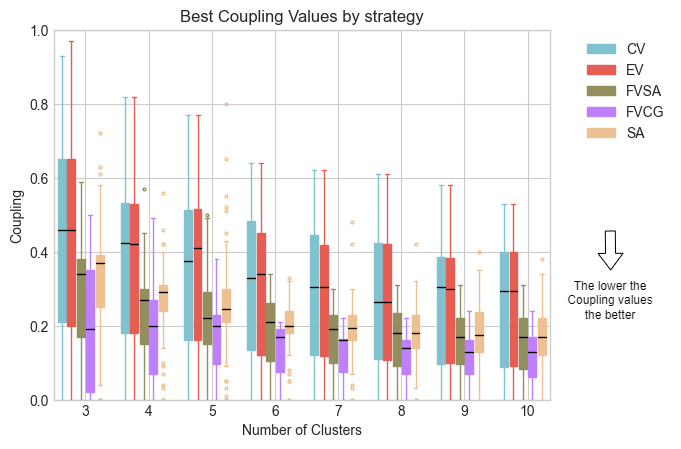}}
\newline
\subfloat[Cohesion\label{fig:best_gen_cohesion}]{\includegraphics[width=0.4\linewidth]{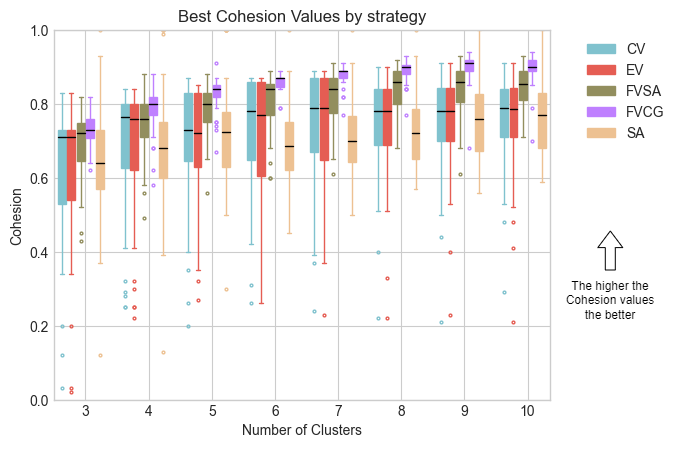}}
\subfloat[Combined\label{fig:best_gen_mix}]{\includegraphics[width=0.4\linewidth]{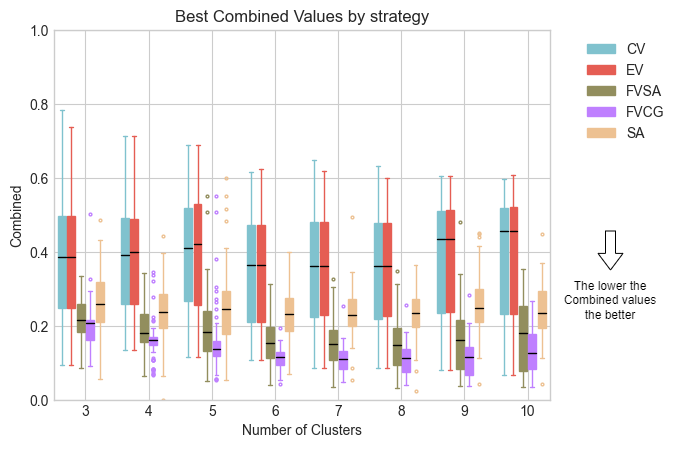}}
\caption{Evaluation Metrics applied to the best decompositions of the 85 codebases for each metric}
\label{fig:best_gen_metrics}
\end{figure*}

Each strategy generated a different number of decompositions (table~\ref{tab:strategies-performance}) derived from the number of parameters and from the conversion of functionalities' clusters to entities' clusters. Note that \textit{FVCG} has the larger number of parameters. Therefore, it was decided to perform a second analysis in which only the best decompositions of each codebase are used for every strategy and number of clusters. This way, the same number of decompositions are considered for each strategy. Additionally, and as is shown in Table~\ref{tab:strategies-performance}, the number of decompositions associated with strategy \textit{FVCG} is significantly larger, which has an impact on the performance. Therefore, it is relevant to understand which parameters can be discarded, if any, to minimize the number of decompositions that need to be generated.

By decreasing the number of decompositions, the results' dispersion of the strategies that generated a higher number of decompositions decreased substantially along with the number of outliers, as shown in Figure~\ref{fig:best_gen_metrics}.

By using only the best decompositions for each codebase, the results of the functionality vectorization strategies improved significantly. For complexity, Welch's t-test accepts the hypothesis that they have the same mean when the number of clusters is 3 and 5. Regarding cohesion, coupling, and the combined metrics, the t-test continues to reject the hypothesis for any number of clusters. Overall, when looking at figure~\ref{fig:best_gen_metrics}, the \textit{FVCG} strategy distinguishes itself from the \textit{FVSA} by having better results for all metrics. Also, the \textit{FVCG} strategy proves to be more interesting because it does not require such an in-depth analysis of the code as the \textit{FVSA} strategy, and is more independent of the technology stack.

After this analysis, we answer each one of the three research questions.    

\subsubsection{Does the use of \textit{Code2Vec} with the functionality perspective provides better results than sequences of accesses?}

To answer the first research question, the proposed strategies that rely on feature vectorization and the \textit{Code2Vec} model (\textit{FVSA}, \textit{FVCG}) are compared with the \textit{SA} strategy, which clusters entities by their access sequences. By comparing the \textit{FVSA} strategy with \textit{SA}, it will be possible to conclude the impact of the \textit{Code2Vec} model on the sequence of entity accesses. And the comparison of \textit{FVCG} and \textit{SA} strategies will indicate if only the use of \textit{Code2Vec} can achieve better results than a very detailed analysis used in the \textit{SA} strategy.

In terms of complexity, Welch's t-test only accepts the hypothesis of 2 strategies having the same mean values when comparing the \textit{FVSA} and \textit{SA} strategies and the number of clusters is 4 or 5. In all other cases, including the \textit{FVCG} strategy, as shown in Figure~\ref{fig:best_gen_complexity}, most of the values of the proposed strategies are lower than those of the \textit{SA} strategy, which leads to the conclusion that using the \textit{Code2Vec} model with a functionality perspective generates less complex decompositions.

Regarding coupling, Figure~\ref{fig:best_gen_coupling} the \textit{FVSA} and the \textit{SA} strategy have very similar results, which can be validated with the results of Welch’s t-test, that accepts the hypothesis that the strategies have the same average coupling values for every number of clusters except for 3 and 5. The \textit{FVCG} strategy obtains better results than the \textit{SA} strategy since the Welch's t-test rejects that both strategies have the same mean for every number of clusters and the \textit{FVCG} coupling results are lower than the ones of the \textit{SA} strategy.

When it comes to the cohesiveness of the proposed strategies~\ref{fig:best_gen_cohesion}, the values are better compared to the \textit{SA} strategy. Welch's t-test rejects all the hypotheses that the \textit{FVCG} and the \textit{FVSA} strategies have the same mean cohesion values when compared to \textit{SA} strategy. This implies that the decompositions generated by the \textit{Code2Vec} proposed strategies have higher cohesive microservices than by using \textit{SA} strategy.

Overall, when applying the combined metric (Figure~\ref{fig:best_gen_mix}) to these strategies, the results of Welch's t-test also reject the hypothesis that the strategies have the same mean values for every comparison between the proposed strategies (\textit{FVCG} and \textit{FVSA}) and the \textit{SA} strategy. With these results, it is possible to conclude that the use of the \textit{Code2Vec} model with a functionality perspective to the sequence of accesses analysis improves the results, but when using just the functionalities vectorization without the sequence of accesses it is possible to achieve even better results.

\subsubsection{Does the application of the functionality perspective provides better results than Al-Debagy and Martinek's class perpective?}

To answer the second research question, the proposed strategies (\textit{FVSA}, \textit{FVCG}) are compared with the \textit{CV} strategy, proposed by Al-Debagy and Martinek, and the \textit{EV} strategy, which is an adaptation of the \textit{CV}.

Calculating Welch's t-test between the proposed strategies and the \textit{CV} strategy it is possible to reject the hypothesis of having the same complexity, cohesion, coupling, and combined means for every number of clusters. These results show that these strategies are quite different as can be seen in figure~\ref{fig:best_gen_metrics} and that the results of the \textit{CV} strategy are worse for every metric than the ones of the \textit{FVCG} and \textit{FVSA} strategies.

\begin{figure}[ht]
\centering
\includegraphics[width=\linewidth]{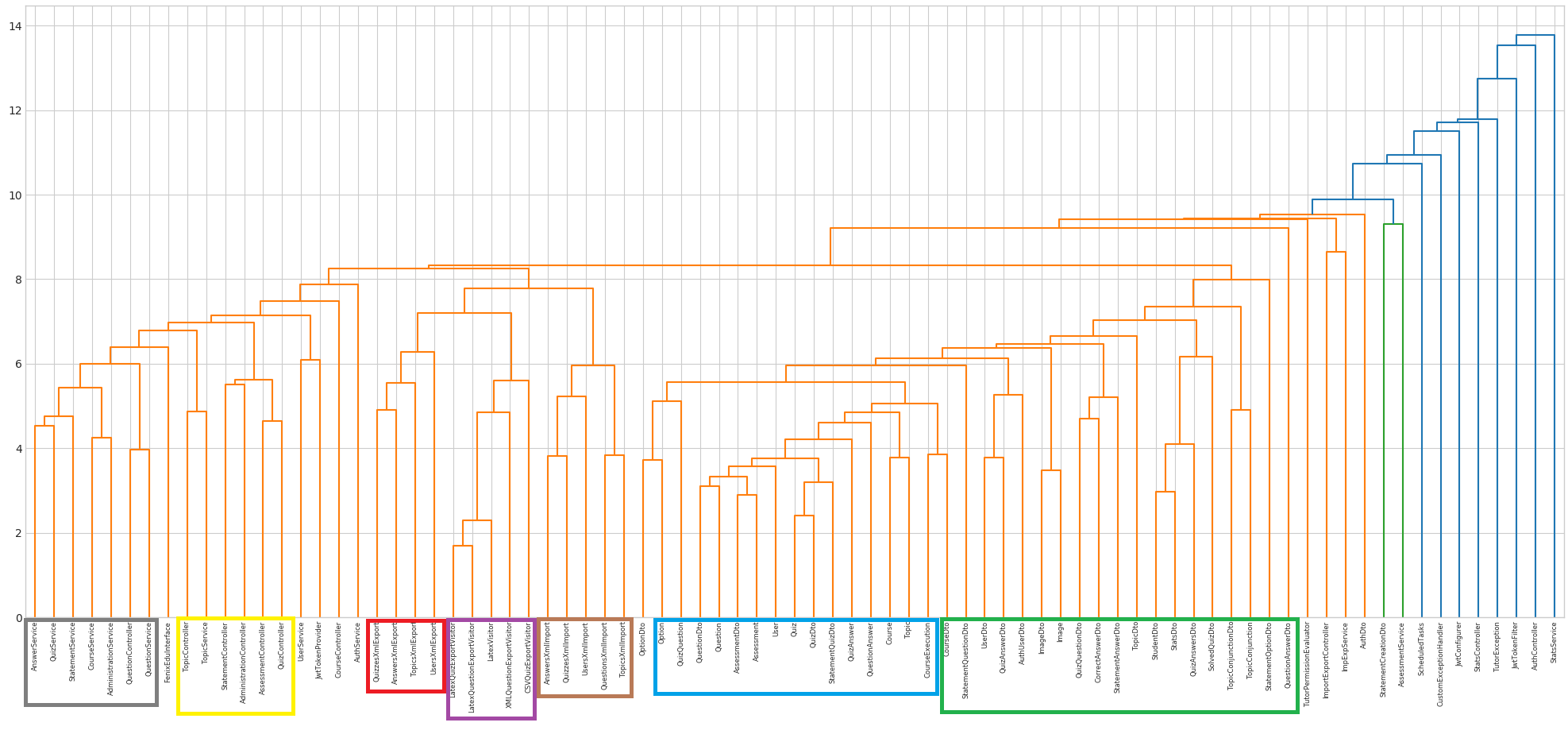}
\caption{Class dendrogram of a codebase using the CV strategy.}
\label{fig:cv-qt-dendrogram-marked}
\end{figure}

As can be observed in the dendrogram in Figure~\ref{fig:cv-qt-dendrogram-marked}, the results of the \textit{CV} strategy derive from clustering the vectorized classes, groups them by their types: Entities, Service, Controller, Configuration. Thus, when class clusters are converted to entity clusters, most of the domain entities are grouped in the same cluster leading to decompositions that have a large cluster.

Trying to avoid this behavior, the \textit{EV} strategy was implemented, which only considers classes that represent domain entities. But, this strategy ended up getting the same results as the \textit{CV} strategy, since Welch's t-test accepts that it has the same average across all strategies and for all numbers of clusters, which led to Welch's t-test also rejecting the hypothesis that this strategy has the same means as the proposed strategies across all metrics and for all numbers of clusters.

This evaluation shows that the use of \textit{Code2Vec} with a class perspective generates worse decompositions, since the class vectors are heavily influenced by each class type because of its respective lexical tokens, but may be a good approach to cluster classes into packages to organize the code by classes types like in~\cite{Hammad2021}.

\subsubsection{Does the input parameters of the proposed strategies impact the results of the evaluation metrics?}

To answer the third research question we analyze the parameters of each of the proposed strategies.

Starting with the \textit{FVCG} strategy, there are six parameters to analyze, the maximum depth (d) the call graph is explored, the four weights to control which method types are more relevant, and the linkage type used in the clustering algorithm.

\begin{figure}[ht]
\centering
\includegraphics[width=1\linewidth]{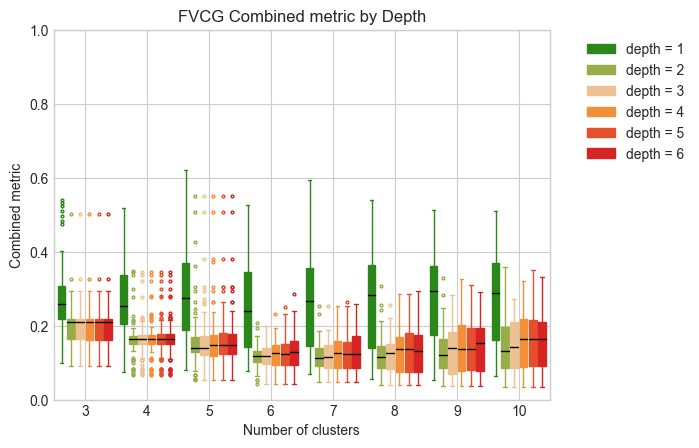}
\caption{Regression of the depth parameter for the combined metric values.}
\label{fig:fvcg-md}
\end{figure}

Figure~\ref{fig:fvcg-md} shows the results for the variation of the depth. Welch's t-test between a depth of 1 and a depth of 2 shows that there is a significant difference between the results, and so depth 1 provides worse results than higher depths. But, when calculating an \textit{OLS} regression for the depths higher than 1, allows us to reject the hypotheses that by increasing the depth more than 2 better results are obtained because the \textit{p-value} is smaller than the significance level.

Therefore, it is impossible to conclude, for depths greater than 1, that any given depth is better than another. Since smaller depths require less computation, it is possible to rely just on depth 2, in which only the functionality controller method and the methods it invokes there are used.

This leads us to conclude that when using a lexical approach with functionalities call graph, only the first methods of each functionality are needed since they present most of the lexical tokens present in the entire functionality call graph.

On the other hand, the regression between the method type weights for the combined metric rejects the hypothesis that a different combination of the method type weights affects the evaluation metric results since the \textit{p-value} is less than the significance level. 

\begin{figure}[ht]
\centering
\includegraphics[width=1\linewidth]{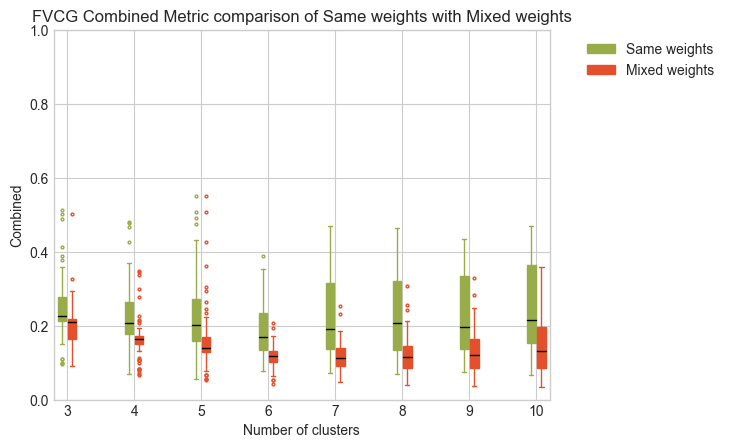}
\caption{FVCG Comparison of the best decompositions combined metric results when the weights are equally distributed versus the best decompositions when using all possible weights distributions.}
\label{fig:fvcg-sw_mw}
\end{figure}

As it is not possible to find a perfect combination of method type weights, we did an additional analysis of the best decompositions for each codebase and number of clusters. This allow us to understand whether the use of the same weights for all method types can achieve good results when compared with all possible weights combinations (Figure~\ref{fig:fvcg-sw_mw}). But, Welch's t-test results reject the hypothesis that the combined metric mean values using the same value for each weight produce better results than a particular weight combination, when looking for the best decomposition.

Regarding the linkage criteria, to understand the impact of the cluster algorithm parameter over the combined metric, it was used a depth of two when comparing the best results for each linkage type: average, simple and complete. 

By comparing the different linkage types for each number of clusters, Welch's t-test allows us to state that the results of both three linkage types have the same means for the combined metric, with just two exceptions where the number of clusters is 6 and 7 when comparing the single type versus the complete linkage type, but even those \textit{p-values} are close to the significance level. This indicates, that the choice of the linkage type is irrelevant when clustering the functionality vectors generated with the \textit{FVCG} strategy.

% === FVSA ===
For the \textit{FVSA} strategy, there are only three parameters to analyze, the two types of access weights, write and read, and also the linkage type.

% Method Type Weights

The regression between the accesses types weights for the combined metric allows us to reject the hypothesis that a different combination of the accesses types weights affects results since the \textit{p-value} is less than the significance level. Once again, as these weights depend on each other, to avoid the problem of multicollinearity, two regressions were made separately, one with just the weights of read accesses, and one with the weights of write accesses. Both regressions reject the hypothesis that the weights have any statistically significant impact on the combined metric results.

% Linkage type

When it comes to the linkage criteria, Welch's t-test rejects the hypothesis that the single type has the same mean as the average type for all cluster sizes except for 3. It accepts the hypothesis that the complete and average types when the number of clusters is 5, 7, 8, 9, and 10, which indicates that for these two linkage types the results obtained are very similar but worse than the single linkage type, which is the one that obtains, in general, the best results for the \textit{FVSA} strategy.

% =============== THREATS TO VALIDITY  ===============
\subsection{Threats to Validity}

% 1 - se a metodologia é generalizável para além de java

The \textit{FVCG} strategy was only implemented to support java codebases, but since it is possible to change \textit{Code2Vec} to accept more languages, it can be easily generalizable, just by creating a new parser for each language.  The parser only needs to generate the methods ASTs.

Due to the codebases selection process, we believe that the 85 selected codebases are representative of monolith systems. Although all codebases use the Spring framework, it does not bias the results, because the frameworks used to develop web monoliths implement the same architectural patterns.

It is possible that there is some correlation between coupling and complexity metrics, so the results of the proposed new combined metric may be biased. Nevertheless, the results are still promising when analyzing each metric separately.

The conversion of functionality clusters to entity clusters may bias the results. However, the strategies that applied this conversion have shown better results.

\subsection{Lessons Learned}

\begin{itemize}
    \item It is possible to perform a lexical analysis of the \textit{AST} with a neural network model and obtain better results than a complex static analysis that captures the functionalities sequences of access to domain entities.
    \item Adding a neural network model to the static analysis of entity accesses (\textit{FVSA} strategy) improves its results.
    \item Classes vectorization is shown to lead to decompositions where the classes are grouped by their type.
    \item The \textit{FVCG} strategy is shown to provide the best results, when compared with the sequence of accesses strategy, and it is only necessary to apply a depth of 2 in the call graph generation, which dramatically improves performance.
\end{itemize}

% ########################################################
% #################### CONCLUSION  #######################
% ########################################################
\section{Conclusion}
\label{sec:conclusion}

As the majority of monolith decomposition approaches perform a static analysis of the source code followed by a clustering algorithm, this work aimed to simplify and generalize this process by recurring to a lexical analysis independent of the technology stack.

The \textit{Code2Vec} model was used to understand that a simple lexical analysis strategy can overcome in terms of complexity, coupling, and cohesion, a more complex analysis that has to extract the functionalities domain entities accesses sequences.

Analyzing monoliths as a set of functionalities was shown to provide better results than the monolith class vectorization strategy, which led to clusters of classes of the same type.

We conclude that the \textit{FVCG} strategy, which only relies on the call graph for functionality vectorization, provides the best results. Additionally, it is possible to reduce the number of parameter combinations to choose the best decomposition by only using depth 2 for the call graph generation.

The approach code and the experimental results are publicly available\footnote{https://github.com/socialsoftware/mono2micro/tree/feature/code2vec}.

\section*{Acknowledgment}
This work was partially supported by Fundação para a Ciência e Tecnologia (FCT) through projects UIDB/50021/2020 (INESC-ID) and PTDC/CCI-COM/2156/2021 (DACOMICO).

\bibliographystyle{./bibliography/IEEEtran}
\bibliography{./bibliography/bibliography}

\end{document}